\documentclass{JHEP3}
\usepackage{amsmath,amsfonts,bbm,euscript}
\usepackage{cite}
\usepackage{epsfig}

\newcommand{\h}[1]{h(r)^{#1}}

\newcommand{\mb}[1]{\mathbf{#1}}

\renewcommand{\bar}[1]{
  \overline{#1}
}

\def\ba{\begin{eqnarray}}
\def\ea{\end{eqnarray}}
\def\be{\begin{equation}}
\def\ee{\end{equation}}     
\def\Str{\text{Tr}^*\;}
\def\Tr{\text{Tr}\;}
\def\det{\text{det}}
\def\ap{{\alpha^{\prime}}}
\def\hf{\frac12}
\DeclareMathOperator{\F}{\EuScript{F}}
\def\G{G}
\def\X{\mathcal{X}}
\def\Y{\mathcal{Y}}
\def\DD{\text{D}\bar{\text{D}}}
\def\lag{\mathcal{L}}

\title{An Improved Brane Anti-Brane Action from Boundary Superstring
  Field Theory and Multi-Vortex Solutions}
\author{Nicholas T. Jones and S.-H. Henry Tye\\
  Laboratory for Elementary-Particle Physics, Cornell University,
  Ithaca, NY 14853.\\
  E-mail: \email{nick.jones@cornell.edu},
  \email{tye@mail.lns.cornell.edu}
}

\abstract{ We present an improved effective action for the
  D-brane-anti-D-brane system obtained from boundary superstring field
  theory.  Although the action looks highly non-trivial, it has simple
  explicit multi-vortex (\emph{i.e.}~codimension-2 multi-BPS D-brane)
  multi-anti-vortex solutions.  The solutions have a curious
  degeneracy corresponding to different ``magnetic'' fluxes at the
  core of each vortex.  We also generalize the brane anti-brane
  effective action that is suitable for the study of the inflationary
  scenario and the production of defects in the early universe.  We
  show that when a brane and anti-brane are distantly separated,
  although the system is classically stable it can decay via quantum
  tunneling through the barrier.}


\preprint{CLNS 02/1807}

\begin{document}

\section{Introduction}
D-branes play a crucial role in string theory
\cite{Polchinski:1998rq}.  To understand the D-brane anti-D-brane
($\DD$) system involves off-shell physics.  A powerful way to study it
is to write down its effective space-time action from
background-independent or boundary string field theory (BSFT)
\cite{Witten:1992qy, Shatashvili:1993kk, Gerasimov:2000zp}. Following
the work on the non-BPS D-brane effective action in open superstring
theory \cite{Kutasov:2000aq}, this program was carried out by two
groups (KL \cite{Kraus:2000nj} and TTU \cite{Takayanagi:2000rz}). Here
we seek to improve on their effective $\DD$ action and study its
properties.

The effective action in Refs\cite{Kraus:2000nj, Takayanagi:2000rz} has
a number of interesting properties. It includes all powers of the
single derivative of the tachyon field $T$, a feature very important
for time dependent, or rolling tachyon, solutions \cite{Sen:2002nu,
Sugimoto:2002fp}.  This feature is also necessary to lead to the fact
that the lower dimensional branes appear as soliton solutions in
tachyon condensation.  In particular, KL/TTU find a codimension-two
BPS brane as a solitonic solution, with the correct brane tension and
the correct RR charge \cite{Sen:1999mg, Witten:1998cd}.
However, that vortex solution does not have ``magnetic'' flux inside
it, contrary to our intuition from the Abelian Higgs model.  As
written, the KL/TTU effective action that involves all powers of the
first derivative of $T$ does not respect the $U(1)\times U(1)$ gauge
symmetry of the $\DD$ system; the derivatives of $T$ do not generalize
to covariant derivatives, as is necessary since the complex tachyon
field $T$ is charged under the relative $U(1)$.  Without the correct
gauge covariant action, it is not clear whether the vortex solution,
and more generally the multi-vortex solutions, should have
``magnetic'' flux inside them or not.

We improve the $\DD$ effective action by restoring the covariance and
the $U(1)\times U(1)$ gauge symmetry of the system so the tachyon
field couples to one of the gauge fields as expected. This improved
action is summarized in Eq.(\ref{action}).  Starting with this $\DD$
action we find analytic multi-vortex multi-anti-vortex solutions (all
parallel with arbitrary positions and constant velocities), summarized
in Eq.(\ref{general_solution}). The solution with $n$ vortices
(\emph{i.e.}~$n$ parallel codimension-2 branes) and $m$ anti-vortices
has total tension $\varepsilon_{p-2} = (n+m) \tau_{p-2}$ and
Ramond-Ramond (RR) charge $\mu_{p-2} = (n-m) \tau_{p-2}g_s$ under the
spacetime $(p-1)$-form potential.  Here $\tau_{p-2}$ is the
D$(p-2)$-brane tension and $g_s$ is the string coupling constant.  The
simplicity of the solution suggests that the $\DD$ effective action
may be useful to study the brane dynamics.  For $m=0$ and an
appropriate choice of the magnetic flux, the solution is
supersymmetric and corresponds to $n$ BPS D$(p-2)$-branes.

These solutions have a curious degeneracy.  Each unit of winding
(\emph{i.e.}~a vortex corresponding to a D-brane) can have up to one
unit of ``magnetic'' flux inside it.  That is, both the tension and
the RR charge are independent of the presence (or absence) of the
``magnetic'' flux.  We expect this degeneracy to be lifted by the
quantum corrections to the $\DD$ action and/or the corrections from
the higher derivative and gauge field-strength terms.  However, it is
not clear exactly how the degeneracy will be lifted.

One motivation to understand the $\DD$ system better is its role in
cosmology. D-brane interaction in the brane world scenario provides a
natural setting for an inflationary epoch in the early universe
\cite{Dvali:1998pa, Burgess:2001fx, Garcia-Bellido:2001ky,
Jones:2002cv, Herdeiro:2001zb} (see also \cite{Quevedo:2002xw} for a
review and extensive list of references).  There, the inflaton is
simply the brane-brane separation while the inflaton potential comes
from their interaction. The simplest such scenario involves a
brane-anti-brane pair\cite{Dvali:2001fw, Burgess:2001fx}. Toward the
end of inflation, as the brane and the anti-brane approach each other
and collide, a tachyon emerges and tachyon condensation
(\emph{i.e.}~the tachyon field rolling down its potential) is expected
to reheat the universe and produce solitons (even codimensional
branes) that appear as cosmic strings in our universe
\cite{Jones:2002cv}. The cosmic string density is estimated to be
compatible with present day observations, but will be critically
tested by cosmic microwave background radiation and gravitational wave
detectors in the near future\cite{Sarangi:2002yt}.  To study inflation
and how it ends, we also construct the $(\DD)_p$ effective action when
the brane is separated from the anti-brane.  We find a separation
dependent tachyon potential which predicts that the $\DD$ system is
classically stable when the brane and anti-brane are further than
$\frac{2\pi^2\ap}{2\ln2}$ apart, but can quantum mechanically decay
with the tachyon tunneling through its potential.  The critical
separation agrees with the result known from other methods
\cite{Banks:1995ch} aside from the factor of $2\ln2$.

The paper is organized as follows. In \S2, we briefly review the BSFT
derivation of the $\DD$ action. Then we use Lorentz and gauge symmetry
to complete the terms in the effective action.  As a check, we expand
it to next to leading order and show agreement with known results.  In
\S3, we present the general multi-vortex multi-anti-vortex solutions,
with zero and non-zero gauge field strengths inside the vortices.  We
calculate the RR charge and the total energy of these solutions and
reveal the degeneracy.  We discuss how this degeneracy may be lifted.
In \S4, we construct the effective action when the D$p$-brane and the
$\bar{\text{D}}p$-brane are separated.  The barrier potential to
tunelling is evaluated.  \S5 is the conclusion.

\section{Brane Anti-Brane Effective Actions}
\subsection{Linear Tachyon Action from BSFT}
We summarize the brane anti-brane effective action from BSFT
calculated by KL and TTU \cite{Kraus:2000nj,Takayanagi:2000rz}.  We
restrict attention to D9-branes in type IIB theory, and generalize
using T-Duality later.  BSFT essentially extends the sigma-model
approach to string theory\cite{Tseytlin:1989rr}, in that (under
certain conditions \cite{Witten:1992qy, Gerasimov:2000zp}) the disc
world-sheet partition function with appropriate boundary insertions
gives the classical spacetime action. This framework for the bosonic
BSFT was extended to the open superstring in \cite{Kutasov:2000aq} and
formally justified in \cite{Marino:2001qc}.  In the NS sector the
spacetime action is
\begin{align}\label{definition_S}
  S_{\text{spacetime}} &= -\int \mathcal DX\mathcal D\psi
    \mathcal D\tilde\psi\;e^{-S_\Sigma-S_{\partial\Sigma}}.
\end{align}
where $\Sigma$ is the worldsheet disc and $\partial\Sigma$ is its
boundary.  The worldsheet bulk disc action is the usual one
\begin{align*}
  S_\Sigma &= \frac1{2\pi\ap}\int d^2z\;
  \partial X^\mu\bar\partial X_\mu 
  + \frac1{4\pi}\int d^2z\left(\psi^\mu\bar\partial\psi_\mu 
  + \tilde\psi^\mu\partial\tilde\psi_\mu\right)\\
  &= \hf\sum_{n=1}^\infty nX_{-n}^\mu X_{n\;\mu} + 
  i\sum_{r=\hf}^\infty\psi_{-r}^\mu\psi_{r\;\mu},
\end{align*}
after expanding the fields in the standard modes.  To reproduce the
Dirac-Born-Infeld (DBI) action for a single brane, the appropriate
boundary insertion is the boundary pullback of the $U(1)$ gauge
superfield to which the open string ends couple; for the $N$ brane $M$
anti-brane system, the string ends couple to the superconnection
\cite{Quillen, Witten:1998cd}, hence the boundary insertion should be
\begin{align}\label{boundary_insertion}
  e^{-S_{\partial\Sigma}} &= \Tr\hat P\exp\left[\int d\tau d\theta 
    \mathcal M(\mb X)\right],&
  \mathcal M(\mb X) &= \left(\begin{array}{cc}
    iA^1_\mu(\mb X)D\mb X^\mu&\sqrt\ap T^\dagger(\mb X)\\
    \sqrt\ap T(\mb X)&iA^2_\mu(\mb X)D\mb X^\mu
  \end{array}\right)
\end{align}
where the insertion must be supersymmetrically path ordered to
preserve supersymmetry and gauge invariance.  $A^{1,2}$ are the $U(N)$
and $U(M)$ connections, and $T$ is the tachyon matrix transforming in
the $(N,\bar M)$ of $U(N)\times U(M)$.  The lowest component of
$\mathcal M$ is proportional to the superconnection.  To proceed, it
is simplest to perform the path-ordered trace by introducing boundary
fermion superfields \cite{Marcus:1987cm}; we refer the reader to
\cite{Kraus:2000nj} for details.  The insertion
(\ref{boundary_insertion}) can then be simplified to be
\begin{align}\label{NM_insertion}
  \Tr P\exp\left[i\ap\int d\tau
    \left(\begin{array}{cc}
      F^1_{\mu\nu}\psi^\mu\psi^\nu+iT^\dagger T 
      + \frac1\ap A^1_\mu\dot X^\mu&
      -iD_\mu T^\dagger\psi^\mu\\
      -iD_\mu T\psi^\mu&
      F^2_{\mu\nu}\psi^\mu\psi^\nu+iTT^\dagger + 
      \frac1\ap A^2_\mu\dot X^\mu
    \end{array}\right)\right],
\end{align}
where the tachyon covariant derivatives are
\begin{align}
  D_{\mu}T = \partial_{\mu} T + iA^1_{\mu}T-iTA^2_{\mu}.\label{T_cderiv}
\end{align}
This expression reproduces the expected results when
the tachyon and its derivatives vanish: the only open string
excitations will be the gauge fields on the branes and the
anti-branes, for each of which the action is the standard DBI action.
For instance, with $N=M=1$, $DT=T=0$, the partition function
(\ref{definition_S}) with the insertion (\ref{NM_insertion}) leads to
\begin{align}\label{F_only}
  S_{\DD} &= -\tau_9\int d^{10}x\left[\sqrt{-\det(g+2\pi\ap F^1)}
    +\sqrt{-\det(g+2\pi\ap F^2)}\right].
\end{align}
The measure in (\ref{definition_S}) was defined to reproduce the
correct tension for the D9-branes, $\tau_9 = 1/[(2\pi)^9 g_s
\ap^5]$\footnote{Throughout this work we assume the dilaton is
  stabilized to give an effective string coupling $e^\phi = g_s$.}.
Unfortunately (\ref{NM_insertion}) cannot in general be simplified,
but for a single brane anti-brane pair, $N=M=1$, demanding that the
gauge field to which the tachyon couples vanishes, $A^-\equiv
A^1-A^2=0$, the path-ordered trace can be performed using worldsheet
boundary fermions.  Writing $A^+=A^1+A^2$, we have \cite{Kraus:2000nj}
\begin{align}\label{1DD_insertion}
  S_{\partial\Sigma} &= -\int d\tau\left[\ap T\bar T + 
      \ap^2(\psi^\mu\partial_\mu T)\frac1{\partial_\tau}
      (\psi^\nu\partial_\nu\bar T) + \frac i2\left(\dot X^\mu A^+_\mu
      +\hf\ap F^+_{\mu\nu}\psi^\mu\psi^\nu\right)\right].
\end{align}
The operator $1/\partial_\tau$ acting on a function $f(\tau)$ is
defined to be the convolution of $f$ with sgn$(\tau)$ over the
worldsheet boundary.  For linear tachyon profiles, gauge and spacetime
rotations allow us to write $T=u_1X^1+iu_2X^2$, and
(\ref{definition_S}) can be calculated, since the functional integrals
are all Gaussian.  The result when $A^+=0$
is derived in \cite{Kraus:2000nj, Takayanagi:2000rz}:
\begin{align}\label{noncov_action}
  S_{\DD} = -2\tau_9\int d^{10}X_0\;\exp&\left[
    -2\pi\ap[(u_1X_0^1)^2+(u_2X_0^2)^2]\right]\F(4\pi\ap^2 u_1^2)
  \F(4\pi\ap^2 u_2^2).
\end{align}
where the function $\F(x)$ is given by\cite{Kutasov:2000aq} 
\begin{align}\label{F_definition}
  \F(x) = \frac{4^xx\Gamma(x)^2}{2\Gamma(2x)}
  = \frac{\sqrt{\pi}\Gamma(1+x)}{\Gamma(\hf+x)}.
\end{align}
Note that $\F(x)=0$ at $x=-1/2$, and 
\begin{align}  
\F(x) = \begin{cases}
    1 + (2\ln2)x + \left[{2(\ln2)^2-\frac{\pi^2}6}\right]x^2
    +\mathcal O(x^3),&
    0<x\ll1,\\\label{F_series}
    \sqrt{\pi x}\left[1 + \frac1{8x} + \mathcal 
  O(\frac1{x^2})\right],& x\gg1, \\
        -1/(1+x), & x \rightarrow -1.
  \end{cases}
\end{align}
This action exhibits all the intricate properties of the $\DD$
system expected from Sen's conjectures: the tachyon potential at its
minima $T\to\infty$ completely cancels the brane tensions; even
codimension solitons can appear on the D9-brane worldvolume, with
exactly the correct tension to be lower dimensional D-branes; odd
codimension solitons on which tachyonic fields reside can appear, with
exactly the tension of the unstable non-BPS branes of type II string
theories \cite{Sen:1999mg}.

BSFT can also give the analogue of the D-brane Chern-Simons action for
the $\DD$ system, defined similarly to (\ref{definition_S}), but
with all fermions in the Ramond sector.  The bulk contribution to the
partition sum can be written as the wave-functional
\cite{Kraus:2000nj, Takayanagi:2000rz}
\begin{align*}
  \Psi^{RR}_{\text{bulk}} &\propto\exp\left[
    -\hf\sum_{n=1}^\infty nX_{-n}^\mu X_{n\;\mu} - 
  i\sum_{n=1}^\infty\psi_{-n}^\mu\psi_{n\;\mu}\right]C,\\
  C&=\sum_{\text{odd }p}\frac{(-i)^{\frac{9-p}2}}{(p+1)!}
  C_{\mu_0\cdots\mu_p}\psi_0^{\mu_0}\cdots\psi_0^{\mu_p}.
\end{align*}
The $\psi_0^\mu$ are the zero modes of the Ramond sector fermions, and
$C_{\mu_0\cdots\mu_p}$ are the even RR forms of IIB string theory.  The
normalization of $\Psi$ can be set later by demanding that the correct
brane charge is reproduced.  The Chern-Simons action is then defined
by
\begin{align*}
  S_{\text{CS}} &= \int\mathcal DX\mathcal D\psi\;\Psi^{RR}_{\text{bulk}}
  \Str P e^{-S_{\partial\Sigma}},
\end{align*}
in which the trace given by
\begin{align*}
  \Str O \equiv \text{Tr}\;\left[\left(
    \begin{array}{cc}\mathbbm{1}_{N\times N}&0\\
      0&-\mathbbm{1}_{M\times M}\end{array}\right)O\right]
\end{align*}
results from the periodicity of the worldsheet fermion superfield
which was necessary to implement to the supersymmetric path ordering.
Again $e^{-S_{\partial\Sigma}}$ can be written as
(\ref{NM_insertion}), with Ramond sector fermions.  This expression
can be viewed as a one dimensional supersymmetric partition function
on $\mathbbm S^1$, and because the Ramond sector fermions are
periodic, this is equivalent to $\Tr (-1)^Fe^{-\beta H}$.  By Witten's
argument \cite{Witten:1982df}, only the zero modes contribute to the
partition sum, giving \cite{Kennedy:1999nn, Kraus:2000nj,
Takayanagi:2000rz}
\begin{align}\label{general_CS}
  S_{\text{CS}} &= \tau_9g_s\int C\wedge\Str e^{2\pi\ap i\mathcal F},\\
  \mathcal F &= \left(\begin{array}{cc}
      F^1 + iT^\dagger T & -i(DT)^\dagger\\
      -iD T & F^2 + iTT^\dagger
    \end{array}\right)\nonumber
\end{align}
$\mathcal F$ is the curvature of the superconnection, and as usual,
the fermion zero modes form the basis for the dual vector space and
all forms above are written with $\psi_0^\mu \to dx^\mu$.  This
expression is exact\footnote{As discussed in \cite{Kraus:2000nj}, this
action is exact in $T$ and $A^\pm$ and their derivatives, but has
corrections for non-constant RR forms.}  and although it was derived
for $2^{m-1}$ brane anti-brane pairs in \cite{Kraus:2000nj,
Takayanagi:2000rz} it appears to have the correct properties for the
general $N$ brane $M$ anti-brane case.

As for the action (\ref{noncov_action}), this result affirms Sen's
conjectures in that it exhibits appropriate coupling to the RR 10-form
potential, and the even codimension solitons have the correct
couplings to the relevant RR forms to be identified as lower
dimensional branes.

\subsection{An Improved $\DD$ Action}
As written, the action (\ref{noncov_action}) for a single brane
anti-brane pair does not manifest the necessary gauge covariance, and
this form of the action is valid only for linear tachyon
profiles\footnote{A covariant perturbative action was derived in
\cite{Takayanagi:2000rz} to order $\ap^2$, but we seek covariance of
the complete action, up to higher derivative terms.  }.  We now
generalize the pure tachyon action of KL and TTU.  Note that there are
precisely two independent Lorentz and $U(1)$ invariant expressions in
terms of first derivatives of the complex tachyon $T$
\cite{Takayanagi:2000rz,Hashimoto:2002xe},
\begin{align*}
  \X &\equiv 2\pi\ap^2g^{\mu\nu}\partial_\mu T\partial_\nu\bar T,&
  \Y &\equiv\left(2\pi\ap^2\right)^2
  \Big(g^{\mu\nu}\partial_\mu T\partial_\nu T\Big)
  \Big(g^{\alpha\beta}\partial_\alpha\bar T\partial_\beta\bar T\Big),
\end{align*}
(with the normalizations chosen for convenience).  For the linear
profile $T=u_1x_1+iu_2x_2$, the only translation invariant way to
reexpress $u_{1,2}$ is as $u_{1,2} = \partial_{1,2} T^{1,2}$; then
with $g^{\mu\nu}=\eta^{\mu\nu}$ we can calculate $\X$ and $\Y$,
\begin{align*}
  \X&=2\pi\ap^2(u_1^2 +u_2^2),\\
  \Y&=\left(2\pi\ap^2\right)^2(u_1^2-u_2^2)^2,
\end{align*}
so the arguments of $\F$ in (\ref{noncov_action}) can be written as
\begin{align*}
  4\pi\ap^2u_1^2 &= \X+\sqrt{\Y},\\
  4\pi\ap^2u_2^2 &= \X-\sqrt{\Y}.
\end{align*}
This provides a unique way to covariantize (\ref{noncov_action}) as
\begin{align}\label{no_gauge_action}
  S_{\DD} = -2\tau_9\int d^{10}x\sqrt{-g}\;&e^{-2\pi\ap T\bar T}
  \F(\X+\sqrt \Y)\F(\X-\sqrt \Y),
\end{align}
which reduces to (\ref{noncov_action}) when $T$ is linear in two
spacetime coordinates.  We shall see in \S\ref{sol_no_gauge} that
restoring the spherical and gauge symmetry in this expression allows
us to construct multiple codimension-2 BPS solitons as expected from
the K-theory arguments \cite{Witten:1998cd}.

Further, we can restore the $A^+$ dependence of the action, since
(\ref{1DD_insertion}) remains quadratic when $A^+\ne0$ if $F^+$ is
constant and the partition function (\ref{definition_S}) will be
Gaussian.  A similar calculation was performed for the non-BPS brane
action \cite{Andreev:2000yn}, and borrowing that result gives the
extended tachyon and gauge field action
\begin{align}\label{A+T_action}
  S_{\DD} &= -2\tau_9\int d^{10}x\;e^{-2\pi\ap T\bar T}
  \sqrt{-\G}\;\F(\X+\sqrt \Y)\F(\X-\sqrt \Y),
\end{align}
where now $\G_{\mu\nu}=g_{\mu\nu}+\pi\ap F^+_{\mu\nu}$ forms the
effective metric for the tachyon, as is usual for open string states
in the presence of a gauge connection \cite{Seiberg:1999vs}:
\begin{align*}
  \X &\equiv 2\pi\ap^2\G^{\{\mu\nu\}}\partial_\mu T\partial_\nu\bar T&
  \Y &\equiv\left|2\pi\ap^2\G^{\mu\nu}\partial_\mu
  T\partial_\nu T\right|^2.
\end{align*}
Indices are raised and lowered with respect to $\G$:
$\G^{\mu\nu}\G_{\nu\alpha} = \delta^\mu_\alpha$, and $\G^{\{\mu\nu\}}$
indicates the symmetric part of $\G$; this symmetrization is
necessary to obtain a real action.
\footnote{It is also possible to include a term in $\X$ proportional
to the anti-symmetric part of $\G^{\mu\nu}$, which must have an
imaginary coefficient for the sake of reality.  The coefficient of
such a term is undetermined by our arguments, and shall be unimportant
in our analysis of the action.}
This coupling to $F^+$ can be
confirmed considering that the $\DD$ system reduces to the non-BPS
brane system under the spacetime IIA $\leftrightarrow$ IIB quotient
$(-1)^{F_L}$ \cite{Sen:1999mg}, which in this system is applied by
setting $\bar T=T$, $F^1=F^2$:
\begin{align*}
  S_{\DD} \xrightarrow[A^1=A^2]{\bar T=T} -2\tau_9\int d^{10}x\;
  e^{-2\pi\ap T^2}\sqrt{-\G}\;\F(2\X)\F(0)
  = \sqrt2 S_{\text{nBPS}}.
\end{align*}
The overall normalization of the action must be divided by $\sqrt2$ to
compensate for the extra boundary fermion in the $\DD$ system
which was integrated over, which is superfluous in the non-BPS brane
system.

The action (\ref{A+T_action}) is still incomplete in that
$A^-=A^1-A^2$, the $U(1)$ connection to which the tachyon couples, was
set to zero in its derivation.  We can conjecture the extension to
$A^-\ne0$ based on the following information:
\begin{itemize}
\item Gauge covariance demands that all tachyon derivatives must be
  replaced by covariant derivatives.  $A^-$ cannot appear outside a
  covariant derivative, so (\ref{A+T_action}) with $\partial T \to DT$
  can only suffer corrections for non-constant $A^-$ (and of course,
  the higher $T$ and $A^+$ derivative corrections).
\item (\ref{F_only}) should be reproduced for $T=DT=0$.
\item We expect the gauge connections to appear in the matrix form 
  \begin{align*}
    \left(\begin{array}{cc}\hf F^+&0\\0&\hf F^+\end{array}\right) \to
    \left(\begin{array}{cc}F^1&0\\0&F^2\end{array}\right),
  \end{align*}
  when we restore $F^- = F^1 - F^2 \ne 0$.  We can insert this into
  the action (\ref{A+T_action}) and trace over the $U(2)$ indices.
\end{itemize}
This leads us to the next improvement to (\ref{A+T_action}), 
\begin{align}
  S_{\DD} &= -\tau_9\int d^{10}x\;e^{-2\pi\ap T\bar T}
  \Bigg[\begin{array}{l}
      \sqrt{-\det[\G_1]}\F(\X_1+\sqrt \Y_1)\F(\X_1-\sqrt \Y_1)\\
      +\sqrt{-\det[\G_2]}\F(\X_2+\sqrt \Y_2)\F(\X_2-\sqrt \Y_2)
    \end{array}\Bigg],\label{action}\\\nonumber
  (\G_{\mu\nu})_{1,2} &\equiv \left(g_{\mu\nu} 
  + 2\pi\ap F^{1,2}_{\mu\nu}\right),
  \quad\quad\quad
  \begin{array}{rl}
    \X_{1,2} &\equiv 2\pi\ap^2\G^{\{\mu\nu\}}_{1,2}
    D_\mu TD_\nu\bar T,\\
    \Y_{1,2} &\equiv\left|2\pi\ap^2\G^{\mu\nu}_{1,2}
    D_\mu TD_\nu T\right|^2,
  \end{array}
\end{align}
The tachyon is charged only under $A^-$:
\begin{align*}
  &D_{\mu}T =\partial_{\mu}T+iA^-_{\mu}T,
  &A^\pm_\mu = A^1_\mu \pm A^2_\mu,
\end{align*}
and the function $\F(x)$ is defined in (\ref{F_definition}).  This is
the effective action which shall be studied in this work.  Corrections
to this action will include higher derivative terms in $T$ and
$F^{\pm}$.  Possible terms like $(F^{-})^{n}T\bar T$ may be included in
higher tachyon derivatives since $[D_\mu,D_\nu]=iF_{\mu\nu}$.  Being
non-supersymmetric, there will be quantum corrections to the action as
well.

In the $\ap$ expansion, using (\ref{F_series}), we have
\begin{align*}
  \F(\X+\sqrt \Y)\F(\X-\sqrt \Y) = 1 + 4 (\ln 2) \X + 
  \left[8 (\ln 2)^2 - \frac{\pi^2}{3}\right] \X^2 - \frac{\pi^2}{3} \Y
  + \ldots
\end{align*}
which agrees with the terms that have four powers of the single 
derivative of $T$ calculated in TTU. This provides a non-trivial check
on the above improved action.  Note that the action is invariant under
$\sqrt{\Y} \rightarrow -\sqrt{\Y}$, so in the above Taylor expansion,
only integer powers of $\Y$ appears in the action.

For time-dependent tachyon fields, $T \rightarrow t/(\sqrt{2\pi}\ap)$, we have 
$\X-\sqrt \Y \rightarrow 0$ and $\X+\sqrt \Y = -\dot T^2 \rightarrow -1$,
so $\F(\X+\sqrt \Y)\F(\X-\sqrt \Y) \rightarrow -1/(1 - \dot T^2)$. 
This justifies the approximation used for the rolling tachyon in
Ref.~\cite{Shiu:2002xp}.

\section{Solitons on Brane Anti-Brane Systems}
Here we shall study the solitonic solutions of the improved $\DD$
effective action (\ref{action}).  Since $T$ is complex, the solitonic
solutions will be vortices, corresponding to D7-branes with brane
tension $\tau_7$.  Let $z = x^1 + i x^2$ be the coordinate in the
complex plane transverse to the D7-branes.  The KL/TTU solution for a
single vortex is given by $T = \lim_{u\to\infty}uz$, $A^\pm = 0$.  We
shall discuss solutions for parallel vortices and anti-vortices.  The
$n$ vortices are located at $\{z_i\}_{i=1,\ldots n}$ while the $m$
anti-vortices are located at $\{z_j^{\prime}\}_{j=1,\ldots m}$.  Since
$T$ is uncharged under $A^+ = A^1+A^2$ only solutions with $A^+=0$ are
studied.  We shall consider an ansatz where the energy density
$\varepsilon_7 = (n+m) \tau_7$ while the total RR charge is
$\mu_7=(n-m)\tau_7g_s$.  We find that there are such solutions with
and without an $A^-$ ``magnetic'' flux associated with each winding
number.  In \S3.1, we show that the RR charge is independent of the
gauge field, or magnetic flux. It is a function of the winding (minus
the anti-winding) number only.  In \S3.2, we calculate the energy
density $\varepsilon_7$ for vortices with and without magnetic
flux. The general solution (\ref{general_solution}) can be found at
the end of this subsection. To understand better the properties of the
solutions, we consider the multi-vortex case more closely in \S3.3.
For an appropriate choice of magnetic flux, the multi-vortex solution
is supersymmetric, though the degeneracy still persists.
We note that the solution for multiple D7-branes without gauge flux we
find was first studied by worldsheet methods in \cite{Hori:2000ic},
and the tensions for multi-kink solitons on non-BPS brane worldvolumes
were calculated in \cite{Hashimoto:2001rk}.

\subsection{Ramond-Ramond Charge of Multi-Soliton Solutions}
\label{sol_no_gauge}
We can gain more insight into the form of the solution giving
multi-soliton branes by looking at the Chern-Simons action
(\ref{general_CS}), which is known exactly.  Multi-soliton solutions
can be constructed with trivial gauge fields, just as in the single
soliton case.  In fact, we show that for soliton solutions, the RR
charge is completely independent of the gauge field to which the
tachyon couples.

Starting with Eq.~(\ref{general_CS}), for D7-brane solitons on a single
brane anti-brane pair, we consider only nonzero RR field $C_8$ , and
set to zero the gauge field under which $T$ is inert, $A^+=0$, $F^+=0$:
\begin{align}\label{C8_ne_0}
  S_{\text{CS}} &= \tau_9g_s\int e^{-2\pi\ap T\bar T}(-iC_8)\wedge\left[
    2\pi\ap iF^--(2\pi\ap)^2DT\wedge D\bar T\right].
\end{align}
The coupling to the field strength, $F^-$, is the standard one, giving
the unstable 9-brane system coupling to 7-branes.  The second term
gives the soliton coupling, and the system can decay to solitons with
trivial gauge fields.  For brevity, we can extract the RR charge
$\mu_7$ of the soliton under a $C_8$ which is constant in the plane in
which $T$ condenses
\begin{align}\label{RR_charge}
  &\mu_7 = -i\frac{\tau_7g_s}{2\pi}\int\limits_{\mathbbm{R}^2}
  e^{-2\pi\ap T\bar T}\left[
    iF^--(2\pi\ap)DT\wedge D\bar T\right],
  &\tau_7=4\pi^2\ap\tau_9.
\end{align}
The single D7-brane solution \cite{Kraus:2000nj, Takayanagi:2000rz},
can be written in polar coordinates on $\mathbbm{R}^2$ as $A^\pm=0$,
$F^{1,2}=0$, $T = uz = ure^{i\theta}$:
\begin{align*}
  \mu_7 &= i\frac{\tau_7g_s}{2\pi}
  \int e^{-2\pi\ap u^2r^2}(2\pi\ap)u^2(-2ir) dr\wedge d\theta\\
  & = \tau_7g_s.
\end{align*}
Unlike in the kinetic term, here $u$ can take any real value without
altering the RR charge of the soliton.

We can construct multi-centered soliton solutions, and in general
$\mu_7$ is independent of the gauge field winding about them.  To
prove this, we require only that:
\begin{itemize}
\item $T=0$ at the center of each soliton, and the tachyon fields
  winds about each of these centers.
\item $T\to\infty$ far from the solitons, so that there the tachyon
  potential and hence the D9-brane anti-brane energy density vanishes.
  Away from the solitons, the D9-brane and anti-brane have annihilated
  and the ground state is indistinguishable from the closed string
  vacuum.
\item $A^-$ can wind only about the soliton centers, in analogy to
  vortices in the Abelian Higgs model.
\end{itemize}

We note first that the terms in (\ref{RR_charge}) can be rewritten using
\begin{align*}
  d\left(e^{-2\pi\ap T\bar T} A^- \right)
  &= e^{-2\pi\ap T\bar T}\left[F^- + 2\pi\ap A^- \wedge
    \left(\bar TdT + Td\bar T\right)\right],\\
  d\left(\hf e^{-2\pi\ap T\bar T}\left[\frac{dT}{T} 
    - \frac{d\bar T}{\bar T}\right]\right)
  &= e^{-2\pi\ap T\bar T}\left(2\pi\ap dT\wedge d\bar T
    + \hf d\left[\frac{dT}{T} - \frac{d\bar T}{\bar T}\right]\right).
\end{align*}
The final term is na\"{\i}vely zero, but receives contributions from
the poles in $dT/T-d\bar T/\bar T$ which result from the zeros
or singularities of $T$; if $T$ is just a polynomial in $z$ and $\bar z$, then
this term is just $2\pi\delta^{(2)}(T,\bar T)$, and each zero (soliton)
contributes equally to this $\delta$-function.  The integrand of
(\ref{RR_charge}) is then
\begin{align}\nonumber
  \mu_7 &= i\frac{\tau_7g_s}{2\pi} \int\limits_{\mathbbm{C}}
  \left\{d\left(\hf e^{-2\pi\ap T\bar T}
  \left[\frac{DT}{T}-\frac{D\bar T}{\bar T}\right]\right)
  - \hf e^{-2\pi\ap T\bar T}
  d\left[\frac{dT}{T} - \frac{d\bar T}{\bar T}\right]\right\},\\
  &= -i\frac{\tau_7g_s}{2\pi}\int\limits_{\mathbbm{C}}
  \hf e^{-2\pi\ap T\bar T}d\left[\frac{dT}{T} 
    - \frac{d\bar T}{\bar T}\right].\label{CS_exterior}
\end{align}
The first term in the first line is the total derivative of a one-form
which vanishes at the boundary at infinity, hence its integral
vanishes.  The remaining term (\ref{CS_exterior}) is independent of
the gauge field and essentially counts the zeros and poles of $T$.
Although this expression might not appear to be gauge invariant, 
the number of zeros and singularities of $T$ is a manifestly gauge
invariant quantity.  Hence $\mu_7$ is not only gauge invariant, but
completely independent of $A^-$ and its curvature.  As an example, if
we construct a solution with $n$ holomorphic and $m$ anti-holomorphic 
zeros
\begin{align*}
  T = u\prod_{i=1}^n(z-z_i) \prod_{j=1}^m (\bar z-\bar z_j^\prime),
\end{align*}
then the total RR D7-brane charge is
\begin{align}
  \mu_7 &= \tau_7g_s\int\limits_{\mathbbm C}
  e^{-2\pi\ap T\bar T}\delta^{(2)}(T,\bar T)dT\wedge d\bar T
  = (n-m)\tau_7g_s.\label{CS_D_Dbar}
\end{align}
Physically, every soliton contributes one topological unit
to the total RR charge of the solution as we expect.  More complicated
solutions include tachyon fields which are multiply wound about their
zeros,
\begin{align}\nonumber
  T &= u\prod_{i=1}^N
  \left(\frac{z-z_i}{\bar z-\bar z_i}\right)^{w_i/2}
  (z-z_i)^{l_i/2}(\bar z-\bar z_i)^{l_i/2},
  \tag{$w_i\in\mathbbm Z$, $l_i\in\mathbbm R^+$}\\
  \mu_7 &= -i\frac{\tau_7g_s}{2\pi}\int\limits_{\mathbbm C}
  e^{-2\pi\ap T\bar T}\hf d\left[\sum_{i=1}^Nw_i\left(
    \frac{dz}{z-z_i}-\frac{d\bar z}{\bar z-\bar z_i}\right)\right]
  = \tau_7g_s\sum_{i=1}^Nw_i.\label{mult_wind}
\end{align}
As we shall see, for this solution to be BPS it must be accompanied by
a gauge field which winds about each $\{z_i\}$; one can explicitly
check that the solution for $A^-$ does not provide any contribution to
$\mu_7$.

These calculations reveal that (\ref{CS_exterior}) behaves in an
intuitive manner; holomorphic or ``positively wound'' zeros of $T$
correspond to D7-branes, and contribute one topological unit to
$\mu_7$, whereas anti-holomorphic or ``negatively wound'' zeros of $T$
represent anti-D7-branes, and contribute oppositely to the RR charge,
the sign arising from the antisymmetry of the volume element.  As for
the single soliton case, it is not necessary to take $u\to\infty$ to
get the exact answer; this is not so when we consider the $\DD$ 
action.

\subsection{Multi-Soliton Tensions}
We now turn to the tension or energy density of the solitons,
beginning with (\ref{action}) and setting $F^+$ to zero.  To obtain
the lowest energy solution it is necessary to take $u$, the overall
multiplying constant in the ansatz for $T$, to $\infty$.  On the
worldsheet, this limit corresponds to the infrared conformal limit, or
equivalently to on-shell physics.  In the effective theory, the limit
allows the tension to be calculated exactly, and since we are
searching for solutions representing classical D-branes, which have
zero width, the regions in the plane at which we require that $V(T\bar
T) = 1$ must be points with all other regions having $V(T\bar T)=0$.
Since $V(T\bar T)=\exp[-2\pi\ap T\bar T]$, the potential will be
maximal at the zeros of $T$, and shall vanish elsewhere when
$u\to\infty$.

We seek to calculate the tension of various solitons; the
energy per 7-volume of the action (\ref{action}) is
\begin{align}\nonumber
  \varepsilon_7 &\equiv \int d^2x \frac2{\sqrt{-g}}
  \frac{\delta S_{\DD}}{\delta g^{0\nu}}g^{\nu0}\\
  &= \tau_9\int d^2x\;e^{-2\pi\ap T\bar T}
  \sqrt{\det(\delta+\pi\ap F^-)}\left[\begin{array}{l}
      \F(\X_++\sqrt \Y_+)\F(\X_+-\sqrt \Y_+)\\
      +\F(\X_-+\sqrt \Y_-)\F(\X_--\sqrt \Y_-)
    \end{array}\right],\label{e7}
\end{align}
when we assume $F^+=0$, all fields are time independent, and work
with flat spacetime.  $\X_\pm,\Y_\pm$ are the tachyon derivative terms
containing the open string metrics $\G = g\pm\pi\ap F^-$
respectively.  When $F^-$ has only a $B$-field component in the 
plane of tachyon condensation, (\ref{F_series}) and some simple
manipulations yield
\begin{align*}
  \lim_{DT\to\infty}&\F(\X+\sqrt \Y)\F(\X-\sqrt \Y)
  = \pi\sqrt{\X^2-\Y},\\
  &= 2\pi^2\ap^2\sqrt{
    \left(D_zTD_{\bar z}\bar T-D_{\bar z}TD_z\bar T\right)
    \left((\G^{\{z\bar z\}})^2D_zTD_{\bar z}\bar T
    -(\G^{\{\bar zz\}})^2D_{\bar z}TD_z\bar T\right)}.
\end{align*}
Until this point, few conditions needed to be placed on the form of
the solutions.  Now there are two simplifying constraints we can
impose; $A^-=0$ motivated from the fact that it seems to be possible
to construct sensible soliton solutions without gauge field winding,
in direct contrast to the solitons of standard field theory.
Secondly, the condition $D_{\bar z}T=0$ was found in
\cite{Hori:2000ic} by worldsheet methods to be the condition which
must be satisfied if $\mathcal N=2$ worldsheet supersymmetry and hence
spacetime supersymmetry is to be preserved; configurations of multiple
parallel branes are mutually BPS and must preserve some spacetime
supersymmetry.  We begin by considering examples that satisfy these
conditions and proceed to other cases.

The first example is one satisfying both conditions; assume $T$ is a
holomorphic function with $n$ zeros at the points $\{z_j\}$,
$T=\lim_{u\to\infty} u\prod_{j=1}^n (z-z_j)$.  Then $T$ represents $n$
separated D7-branes, although the result is identical when some
D7-brane locations coincide.  The gauge field is trivial, hence
$\G^{z\bar z} = 2$ and the tension (\ref{e7}) becomes (after taking
$u\to\infty$)
\begin{align*}
  \varepsilon_7 &= 2\tau_9\int e^{-2\pi\ap T\bar T}4\pi^2\ap^2\;
  |\partial_zT\partial_{\bar z}\bar T|\;\frac i2dz\wedge d\bar z,\\
  &= i\tau_9\int e^{-2\pi\ap T\bar T}(2\pi\ap)^2dT\wedge d\bar T.
\end{align*}
Since $\partial_zT\partial_{\bar z}\bar T$ is always positive, the
absolute value could be ignored.  This is identical (up to the factor
of $g_s$) to the D7-brane charge under the RR 8-form field
(\ref{C8_ne_0}); it was shown above that this is always equal to
$n\tau_7g_s$, hence the multi-solitons have the correct tension
($\varepsilon_7=n\tau_7$) and exhibit BPS properties.  Further, we can
formulate solutions with vortices moving at constant velocities, $z_j
= z_{j,0}+v_j t$.  Of course, the second line of (\ref{e7}) is no
longer valid when the solution has time dependence, so the first line
must be used to give the general form when $T$ is time dependent.  The
velocity dependence leads to the special relativistic $\gamma$-factors
in the energy density of the resultant solution,
\begin{align}\label{gamma-factors}
  \varepsilon_7 = \tau_7\sum_{j=1}^n\frac1{\sqrt{1-|v_j|^2}}.
\end{align}
This result relies on the form of the kinetic action; for instance, in
the case where $n=1$, $\X\sim 2-|v_1|^2$ and $\Y\sim |v_1|^4$, so only
the combination $\sqrt{\X^2 -\Y} \sim \sqrt{1-|v_1|^2}$ together with
the variations of $\F(\X\pm\sqrt\Y)$ with respect to $g^{00}$
give the correct dilation factors above.

Another example is condensation to a D7-brane anti-brane pair which is
obviously non-supersymmetric; in this case we still expect tension of
$2\tau_7$, whereas (\ref{CS_D_Dbar}) shows that the total RR charge is
zero.  Placing the soliton and anti-soliton on the real axis at $x_0$
and $-x_0$ respectively $T=u(z-x_0)(\bar z+x_0)$, after taking
$u\to\infty$ we have
\begin{align*}
  \varepsilon_7 &= 2\tau_9\int e^{-2\pi\ap T\bar T} (2\pi\ap)^2
  |\partial_zT\partial_{\bar z}\bar T
  - \partial_{\bar z}T\partial_z\bar T|\;\frac i2dz\wedge d\bar z.
\end{align*}
For $x_0=0$, the brane and anti-brane are coincident, there is no
winding of the tachyon field, and the total tension is zero (the
tachyon derivative terms cancel).  For any $x_0>0$, the absolute value
gives two regions of integration with opposite signs
\begin{align*}
  \varepsilon_7 &= i\tau_9\left(\;\int\limits_{\Re(z)>0} -
  \int\limits_{\Re(z)<0}\right) e^{-2\pi\ap T\bar T}
  (2\pi\ap)^2dT\wedge d\bar T.
\end{align*}
By the arguments of the previous section, the first integral receives
only the positive contribution from the zero of $T$ at $z=x_0$, the
second only the negative contribution from the zero at $\bar z =
-x_0$, giving $2\tau_7$ as the total tension.  Care must be taken
because the arguments of the previous section relied on Stoke's
theorem, and here we are introducing a new boundary along the
imaginary axis, however since we work in the limit $u\to\infty$ and
the boundary integrand is proportional to $\exp[-2\pi\ap T\bar T]$,
the boundary term vanishes, and the result remains valid.

We can use the understanding gained from this experience to calculate
the tension of the configuration $T = \lim_{u\to\infty}u
\prod_{i=1}^n(z-z_i) \prod_{j=1}^m(\bar z-\bar z_j^\prime)$,
representing $n$ D7-branes and $m$ anti-branes (all parallel).  We
assume no brane and anti-brane position coincides $z_i\ne
z_j^\prime,\forall \{i,j\}$.  The tension is
\begin{align*}
  \varepsilon_7 &= 2\tau_9\int e^{-2\pi\ap T\bar T}(2\pi\ap)^2
  T\bar T\left|\sum_{i,k=1}^n\frac1{z-z_i}\frac1{\bar z-\bar z_k}
  - \sum_{j,l=1}^m\frac1{z-z_j^\prime}\frac1{\bar z-\bar z_l^\prime}
  \right|\;\frac i2dz\wedge d\bar z.
\end{align*}
In regions about each $z_i$ or $z_j^\prime$ (D7-brane or D7-anti-brane) 
the term in absolute values is positive or negative respectively.
Denoting these regions by $\Gamma_i$ and $\Gamma_j^\prime$ the tension
is
\begin{align*}
  \varepsilon_7 &= i\tau_9\left(\sum_{i=1}^n\;\int\limits_{\Gamma_i}
  - \sum_{j=1}^m\;\int\limits_{\Gamma_j^\prime}\right)
  e^{-2\pi\ap T\bar T}(2\pi\ap)^2dT\wedge d\bar T.
\end{align*}
Each integral precisely resembles (\ref{C8_ne_0}) and so the value of
the integral is proportional to $+1$ for each holomorphic zero and
$-1$ for each anti-holomorphic zero of $T$ in the region.  The
boundary terms again vanish because we take $u\to\infty$, giving the
expected result
\begin{align*}
  &\varepsilon_7 = (n+m)\tau_7,
  &\mu_7 = (n-m)\tau_7g_s,
\end{align*}
where the result of the RR charge calculated earlier has been included
for comparison.

\subsection{Vortex Tension with Magnetic Flux}
More complicated solutions include gauge field winding about zeros of
the tachyon field.  For such solutions the RR 8-form charge is given
by (\ref{mult_wind}), irrespective of the behavior of the gauge field.
The tachyon fields wind more than once around each zero when $T$ is
not entirely holomorphic.  This represents multiple D7-branes which
preserve $\mathcal N = 1$ spacetime supersymmetries; the necessary conditions
to obtain such BPS configurations are $D_{\bar z}T = 0$ and $F_{zz} =
F_{\bar z\bar z} = 0$\cite{Hori:2000ic}.  The later condition is
trivial in this system, but imposing the former determines the form of
$A^-$ and its curvature
\begin{align}
  T &= \lim_{u\to\infty} u\prod_{i=1}^N
  \left(\frac{z-z_i}{\bar z-\bar z_i}\right)^{w_i/2}
  (z-z_i)^{l_i/2}(\bar z-\bar z_i)^{l_i/2},
  \tag{$w_i\in\mathbbm Z^+$}\\
  A^-_{\bar z} &= -\frac i2\sum_{i=1}^N\frac{a_i}
  {\bar z-\bar z_i},\quad\quad\quad
  F^-_{z\bar z} = -2\pi\sum_{i=1}^Na_i
  \delta^{(2)}(z-z_i,\bar z-\bar z_i),\label{mult_wind_gauge}
\end{align}
For this configuration, since
\begin{align*}
  D_{\bar z}T = \frac T2\sum_{i=1}^N \frac{l_i-(w_i-a_i)}
  {\bar z - \bar z_i},
\end{align*}
we must have $l_i = w_i-a_i, \forall i$ to be a BPS configuration.  In
order to obtain a solution with just one unit of magnetic flux for
each winding number, $w_i = a_i$, $l_i$ must be zero; this can be
achieved by taking the limit $l_i\to0$ in such a way that
$\lim_{l_i\to0,u\to\infty} ul_i \to \infty$.

To calculate the tension of this ansatz, we must apply a
regularization of the $\delta$-functions, and the tension is
regularization dependent; we choose that which gives the tension to be
independent of the gauge field winding, to show that such a solution
is possible.  Formally this requires that we write the
$\delta$-function in $F^-$ as a Gaussian of width $\epsilon$;
requiring that $\frac1\epsilon\sim u^{n>2}$ implies taking
$\epsilon\to0$ before $u\to\infty$, and the tension of the solution
will be equal to the RR charge.  In this regularization, before taking
$u\to\infty$ we split the integral into regions about each zero of $T$
as before, and split each region $\Gamma_i$ into one about the pole at
$z_i$ $(\Gamma_{i,\le\epsilon})$ and one over the rest of the region
$(\Gamma_{i,>\epsilon})$
\begin{align*}
  \varepsilon_7 &= \lim_{u\to\infty}\lim_{\epsilon\to0}
  \sum_{i=1}^N\left(\;\int\limits_{\Gamma_i\le\epsilon}
    +\int\limits_{\Gamma_i>\epsilon}\right)\lag.
\end{align*}
The first integral can be evaluated using 
\begin{align*}
  \sqrt{\det(\delta+\pi\ap F^-)} &= \sqrt{1+\hf(\pi\ap F^-)^2}
  &\longrightarrow&&&4\pi^2\ap\left|a_i\right|
  \delta^{(2)}(z-z_i,\bar z-\bar z_i),\\
  ({\G}^{-1})^{\mu\nu} &= \frac{g^{\mu\nu}-\pi\ap F^{-\mu\nu}}
       {1+\hf(\pi\ap F^-)^2}
  &\longrightarrow&&&0,
\end{align*}
about $z=z_i$, which gives for the contribution to the tension
from $\Gamma_{i,\le\epsilon}$
\begin{align*}
  2\tau_9\int\limits_{\Gamma_{i,\le\epsilon}}
  4\pi^2\ap\left|a_i\right|
  \delta^{(2)}(z-z_i,\bar z-\bar z_i)\frac i2 dz\wedge d\bar z
  = \left|a_i\right|\tau_7.
\end{align*}
The remaining integral is over that part of $\Gamma_i$ further than
$\epsilon$ from $z_i$.  Na\"{\i}vely this should be zero because we
have argued that only the zeros of $T$ contribute to the soliton
tension and charge, but the specific regularization of the solution 
using $\frac1\epsilon\sim u^{n>2}$ gives us $\tau_7|w_i-a_i|$, it being
necessary to take $u\to\infty$ after taking $\epsilon\to0$.  The total
energy density of the soliton is the sum over the integrals in both
regions for all $\Gamma_i$ and is
\begin{align}\label{tension_w_gf}
  \varepsilon_7 &= 
  \tau_7 \sum_{i=1}^N\left(|w_i-a_i|+\left|a_i\right|\right).
\end{align}
Since $w_i$ are positive (allowing them to take negative values would
change some branes to anti-branes), when all $w_i \ge a_i \ge 0$ the
solution has minimal energy and the tension is equivalent to the RR
8-form charge (\ref{mult_wind}), $\varepsilon_7 = \tau_7
\sum_{i=1}^Nw_i = \mu_7/g_s$.  Therefore we appear to have multiple
solutions representing certain brane systems, with different degrees
of gauge field winding but with identical energy and RR charge. At
this level in the effective theory, it is a curious degeneracy of the
soliton solutions, which is likely lifted by higher derivative and/or
quantum corrections to the effective action.

To summarize, we give the general ansatz for a tachyon field
representing a set of parallel $n$ D7-branes and $m$ anti-branes.  The
energy per 7-volume of this solution is $\varepsilon_7=(n+m)\tau_7$
and its RR charge under the spacetime 8-form potential is
$\mu_7=(n-m)\tau_7g_s$.
\begin{align}
  T &= \lim_{u\to\infty}u\prod_{i=1}^N
  \left(\frac{z-z_i}{\bar z-\bar z_i}\right)^{w_i/2}
  |z - z_i|^{l_i}\prod_{j=1}^M
  \left(\frac{\bar z-\bar z_j^\prime}{z-z_j^\prime}\right)^{w_j^\prime/2}
  |z - z_j^\prime|^{l_j^\prime},\nonumber\\
  A^-_{\bar z} &= -\frac i2\sum_{i=1}^N\frac{a_i}
  {\bar z-\bar z_i}
  - \frac i2\sum_{j=1}^M\frac{a_j^\prime}
  {\bar z-\bar z_j^\prime},\label{general_solution}
\end{align}
where $z_i$ ($z_j^\prime$) are the constant positions of the
(anti-)branes.  Single valuedness of $T$ requires that
$\{w_i,w_j^\prime\}$ is some set of positive integers, and we have
defined $\sum_{i=1}^Nw_i=n, \sum_{j=1}^Mw_j^\prime = m$.  $\{a_i,
a_j^\prime\}$ must satisfy $0\le a_i \le w_i$, $0\le a_j^\prime \le
w_j^\prime$ in order to obtain the minimal energy solution as in
(\ref{tension_w_gf}).  If the brane or anti-brane were to move at
constant velocities, the tensions would pick up the relativistic
$\gamma$-factors as in (\ref{gamma-factors}).

\subsection{Discussion}
The richness of vortex solutions examined in this section is
vindication of the gauge covariant form of the tachyon kinetic terms
used; the tensions of all solitons are as expected, the results can be
calculated in any coordinate system, and by persuading the solitons to
move at constant velocities the necessary special relativistic factors
arise.  All this evidence depends crucially on the $\X\pm\sqrt \Y$
structure of the action.

The usual topological arguments suggest the vortices of the action
(\ref{action}) are stable and we verify this by perturbing a
characteristic solution representing $n$ coincident D7-branes,
\begin{align*}
  &T = \lim_{u\to\infty}u\left(\frac z{\bar z}\right)^{n/2}
  \left(z\bar z\right)^{l/2} + t(z,\bar z),
  &A^-_{\bar z} = -\frac{ia}{2\bar z}.
\end{align*}
The first order perturbations vanish for all values of $[l,n,a]$, so
these are solutions of the equations of motion.  When the condition
for $\mathcal N = 2$ worldsheet supersymmetry is satisfied,
\begin{align}\label{BPS_condition}
  &D_{\bar z}T = 0, &\text{or}&& l = n-a,
\end{align}
the second order perturbations are
\begin{multline}
  \frac{\delta^2 S}{\delta T^2}t^2 
  + \frac{\delta^2 S}{\delta \bar T^2}\bar t^2
  + 2\frac{\delta^2 S}{\delta T\delta\bar T}t\bar t \propto
  -le^{-2\pi\ap u^2|z|^{2l}}\left(
  \hf\eta^{ij}\partial_it\,\partial_j\bar t
  + a\delta^{(2)}(z,\bar z)t\bar t\right)\\
  - \underbrace{le^{-2\pi\ap u^2|z|^{2l}}\Big[\overbrace{
      \left(\partial_z t\partial_{\bar z}\bar t
      +\partial_{\bar z}t\partial_z\bar t\right)
      }^{\text{from }\X^2} - \overbrace{
      2\partial_{\bar z}t\partial_z\bar t}^{\text{from }\Y}
  +\ldots\Big]}_{\text{total derivative}}
\end{multline}
The terms other than the kinetic terms in the off-brane directions
conspire to form a total derivative (again because of the form of the
tachyon kinetic terms, $\X\pm\sqrt \Y$) leaving just the modes in the
directions along the D7-branes.  When $a=0$, they represent the two
massless fluctuations of the set of branes in the two transverse
dimensions.  When $a\ne0$ the gauge field must be perturbed similarly
otherwise the fluctuations are massive; checks reveal that the gauge
field perturbations are likewise stable.

Let us recall Derrick's theorem (see \cite{Coleman:1985} for
instance); in a field theory of a set of scalar fields, suppose there
is a time-independent solitonic solution $T(x)$ with codimension
$d_c$.  Consider the one-parameter family of field configurations
defined by $T(x;\lambda) \equiv T(\lambda x)$ where $\lambda$ is
positive and real. In general the energy is $E(\lambda) =
\lambda^{-d_c}P + \lambda^{-d_c+2}K_2 + ... $, where $P$ is the
potential energy contribution (defined so $P \ge 0$) and $K_2$ is the
two derivative kinetic energy term, and extra terms may be present if
there are more than two derivative terms in the theory. By Hamilton's
principle, this must be stable at $\lambda=1$, that is, ignoring
possible extra terms, $(d_c-2)K_2 + d_c P =0$.  Since both $P$ and
$K_2$ are positive in an ordinary field theory, only codimenion
$d_c=1$ soliton is possible. Vortices (codimension two) in Abelian
Higgs model are possible due to the presence of magnetic flux.  Now we
look at vortices in the $\DD$ system.  The energy of a
time-independent soliton $T(x)$ with codimension $d_c$, in the limit
$DT\to\infty$ is $E = 2\pi\tau_9\int d^{(d_c)}x V(T\bar T)
\sqrt{\X^2-\Y}$.  That is $P=0$ which implies $E$ scales like a two
derivative term and
\begin{align*}
  (d_c-2) E = 0
\end{align*}
Therefore only codimension two solitons are possible. In contrast to
the Abelian Higgs model, a magnetic flux is not necessary for the
existence of the vortices in the $\DD$ system, as we have seen.

Returning to the issue of the apparent degeneracy between vortices
with and without $A^-$ flux, we repeat that this degeneracy is
expected to be lifted by corrections to the effective action.  The
effective action has at least two sources of corrections:
\newcounter{lcount}
\begin{list}{{\bf \roman{lcount}.}}{\itemsep -5pt}\usecounter{lcount}
\item Classically there are higher derivative terms.
\item Since the $\DD$ system is non-supersymmetric, there will be
  quantum corrections.
\end{list}
Such corrections should lift the degeneracy.  For any solution that
involves both vortices and anti-vortices, supersymmetry is clearly
broken; a more careful calculation would reveal tachyonic modes in
such systems. Considering only vortices which satisfy the BPS
condition $l=n-a$, the solution with no gauge field ($a=0, l=n$) can
only receive corrections from multiple (anti-)holomorphic derivatives
of $T$ ($\bar T$).  Since $T\sim z^n$, only the first $n$ holomorphic
derivatives of $T$ are non-zero.  Importantly the degeneracy is
already present in the single vortex solution : the $n=1$ solution may
have flux $a=0$ or up to $a=1$, with the same energy and RR charge.
Note that the $n=1$ solution without flux \cite{Kraus:2000nj,
Takayanagi:2000rz} is classically exact: there are no gauge field
derivative corrections and all second and higher derivatives of the
tachyon field vanish.  If the degeneracy is lifted, one may
na\"{\i}vely conclude that this zero-flux solution will be the stable
one.  If this is the exact BPS solution, then putting $n$ of them
together should be an exact solution too. However, higher derivative
terms are non-zero for the $n$-vortex solution ($n>1$, still with
$a=0$).  Barring a miraculous cancellation, the $n$-vortex solution
will have classical corrections from the higher derivative terms.

Another solution to consider is that with $n=a$, or $l=0$ from the BPS
condition (\ref{BPS_condition}).  Then noting that $D_zT =
\frac{(l+(n-a))}{2z}T=\frac{l}{z}T$, for $l=0$ both $D_{\bar z}T =
D_zT = 0$, na\"{\i}vely implying that all higher derivative terms
vanish and $n=a$ is an exact solution.  Recall, however, that the
solution with a wound gauge field requires some care in taking the
limit, and we can only take $l\to0$ with $u\to\infty$ and
$lu\to\infty$, implying $\lim_{l\to0}D_zT \ne 0$.  Therefore
corrections due to higher derivative terms cannot be ignored in the
case where there is a flux associated with each winding,
\emph{i.e.}~$n=a$.  In conclusion, we do not know how the degeneracy
will eventually be lifted.

\section{Lower Dimensional Brane Anti-Brane Systems}
Lower dimensional brane anti-brane pairs can be constructed in a
straightforward manner by applying T-duality to the action
(\ref{action}).  In these systems, the brane and anti-brane can be
separated because under T-duality components of $A^-$ transform into
the relative separation of the pair.  We follow closely the procedure
of \cite{Myers:1999ps}.  Because under T-duality, both the dilaton
transforms and there is mixing between the metric and the Kalb-Ramond
$B$-field, it is necessary to include these in our action
(\ref{action}).  The T-duality properties of the various fields in the
action are well known; the gauge fields in the T-dual directions
transform into the adjoint scalars, the metric and Kalb-Ramond field
mix, the string coupling scales.  Being an open string scalar state,
the tachyon is inert under T-duality.  Under T-duality in directions 
labeled by uppercase Latin indices, (lowercase Latin indices
labeling unaffected directions on the brane), the fields transform as
\cite{Myers:1999ps}
\begin{align*}
  T &\to T,
  &A_a &\to A_a,
  &A_I &\to \frac{\Phi^I}{2\pi\ap},\\
  E_{\mu\nu} &\equiv g_{\mu\nu} + B_{\mu\nu},
  &e^{2\phi} &\to e^{2\phi}\det E^{IJ},
  &E_{IJ} &\to E^{IJ}\\
  E_{ab} &\to E_{ab}-E_{aI}E^{IJ}E_{Jb},
  &E_{aI} &\to E_{aK}E^{KI},
  &E_{Jb} &\to -E^{JK}E_{Kb},
\end{align*}
where $E^{IJ}$ is the matrix inverse to $E_{IJ}$.  The result of
T-dualing $9-p$ dimensions can be written most simply by defining the
pull-back in normal coordinates as:
\begin{align*}
  &P[E_{ab}]^{1,2} \equiv E_{ab} + E_{I\{a}\partial_{b\}}
  \Phi^{I\;1,2} +
  E_{IJ}\left(\partial_a\Phi^I\partial_b\Phi^J\right)^{1,2},
  &P[E_{aI}]^{1,2} &\equiv E_{aI} + E_{JI}\partial_a\Phi^{J\;1,2}.
\end{align*}
Care must be taken because there are two sets of scalars which
describe the position of each brane; they are denoted herein by
$\Phi^{I\;1,2}$ and their difference as $\varphi^{I} \equiv
\Phi^{I\;1} - \Phi^{I\;2}$ which is the scalar representing the
$(\DD)_p$ separation.  In calculating the pull-back of any quantity
only the indices corresponding to directions along the brane are
affected. After T-dualing the fields in (\ref{action}) as above and
performing manipulations similar to those in \cite{Myers:1999ps}, we
obtain the improved action for a D$p$ brane anti-brane pair:
\begin{align}\nonumber
  S_{(\DD)_p} = -\tau_p\int d^{p+1}x\;e^{-2\pi\ap T\bar T}&
  \left[
    \sqrt{-\det[\G_1]}\F(\X_1+\sqrt \Y_1)\F(\X_1-\sqrt \Y_1)\right.\\&+
    \left.\sqrt{-\det[\G_2]}\F(\X_2+\sqrt \Y_2)\F(\X_2-\sqrt \Y_2)
    \right]\label{T-dual_action}
\end{align}
where now the effective metric contains the spacetime metric
pulled-back to the brane worldvolume (and includes any non-zero NS-NS
B field)
\begin{align*}
  &\G^{1,2}_{ab} \equiv P[E_{ab}]^{1,2}+2\pi\ap F^{1,2}_{ab},
\end{align*}
and the covariant derivative dependence of $\X$ and $\Y$ in
(\ref{action}) leads to $\Phi$ dependence in the T-dual action.  The
complete expressions for $\X$ and $\Y$ are
\begin{align*}
  \X_{1,2} &= 2\pi\ap^2
  \left[\begin{array}{r}
      \G_{1,2}^{\{ab\}}D_aTD_b\bar T 
      + \frac1{(2\pi\ap)^2}\varphi^{I}\varphi^{J}T\bar T
      \left(E_{\{IJ\}} -\G_{1,2}^{ab}P[E_{\{Ia}E_{bJ\}}]^{1,2}\right)\\
      +\frac i{2\pi\ap}\left(\G_{1,2}^{ab}P[E_{bI}]^{1,2}
      -\G_{1,2}^{ba}P[E_{Ib}]^{1,2}\right)
      \left(TD_a\bar T-\bar TD_a T\right)\varphi^I
    \end{array}\right]\\
  \Y_{1,2} &= \left(2\pi\ap^2\right)^2
  \left|\begin{array}{r}
    \G_{1,2}^{ab}D_aTD_bT 
    +\frac i{2\pi\ap}\G_{1,2}^{ab}\left(P[E_{bI}]^{1,2}D_aT
    -P[E_{Ia}]^{1,2}D_bT\right)T\varphi^{I}\\
    - \frac1{(2\pi\ap)^2}\varphi^{I}\varphi^{J}T^2
    \left(E_{IJ}-\G_{1,2}^{ab}P[E_{Ia}E_{bJ}]^{1,2}\right)
    \end{array}\right|^2.
\end{align*}
These expressions simplify considerably in Minkowski spacetime when
$B=0$ and $A^{1,2}=0$:
\begin{align*}
  \G_{ab}^{1,2} &= \eta_{ab} + 
  \delta_{IJ}(\partial_a\Phi^I\partial_b\Phi^J)^{1,2},\\
  \X_{1,2} &\xrightarrow[B=0,\;A^{1,2}=0]{g=\eta}\;
  2\pi\ap^2\left[\G_{1,2}^{ab}\partial_aT\partial_b\bar T 
    + \frac1{(2\pi\ap)^2}\varphi^{I}\varphi^{J}T\bar T
    \left(\delta_{IJ} - \G_{1,2}^{ab}
    \partial_a\Phi^{1,2}_I\partial_b\Phi^{1,2}_J\right)\right],\\
  \Y_{1,2} &\xrightarrow[B=0,\;A^{1,2}=0]{g=\eta}\;
  \left(2\pi\ap^2\right)^2\left|\G_{1,2}^{ab}\partial_aT\partial_bT
  - \frac1{(2\pi\ap)^2}\varphi^{I}\varphi^{J}T^2
  \left(\delta_{IJ} - \G_{1,2}^{ab}
  \partial_a\Phi^{1,2}_I\partial_b\Phi^{1,2}_J\right)
  \right|^2.
\end{align*}
It is clear that the action contains relative velocity dependent
terms.  It would be interesting to study the implications of the
velocity dependence of this action to density perturbations in brane
inflationary models.

\FIGURE[r]{\parbox{8cm}{\epsfig{file= 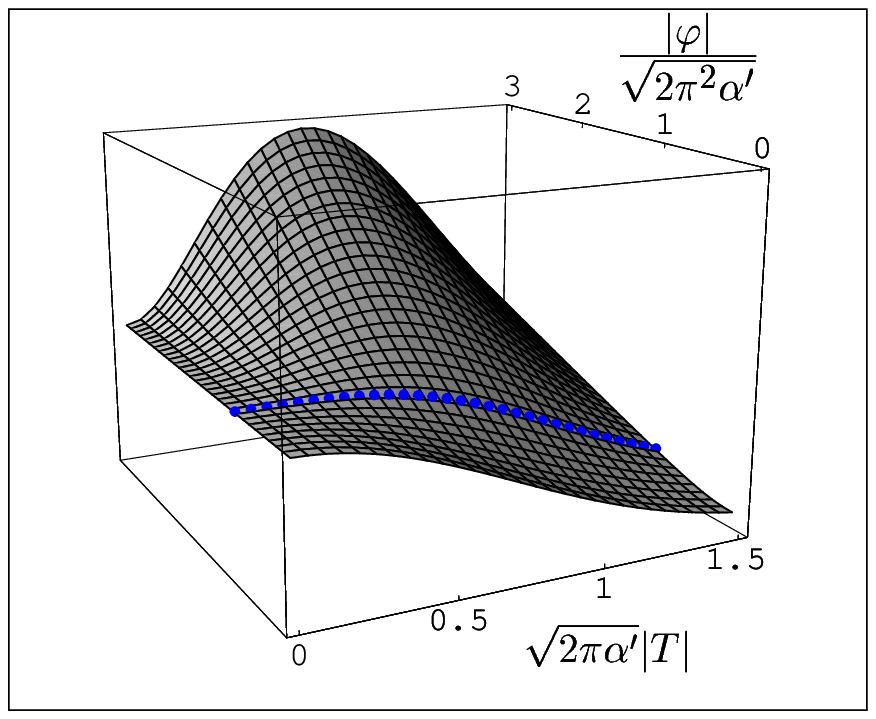,width=8cm}}
  \caption{Separation dependent tachyon potential.  The blue dotted
  line marks the critical separation.}\label{figure_potential}
}
Some important properties of the separation dependent tachyon
potential can be verified. The potential is equal to that part 
of the Lagrangian which is independent of gauge fields and 
derivatives,
\begin{align*}
  &V(T,\varphi) = 2\tau_p e^{-2\pi\ap T\bar T}
  \F\left(\frac1\pi|\varphi|^2T\bar T\right),
  &|\varphi|^2 &\equiv E_{IJ}\varphi^I\varphi^J,
\end{align*}
which gives as the position dependent mass of the tachyon
\begin{align*}
  m_T^2 &= \frac1{2\ap}\left(\frac{|\varphi|^2}{2\pi^2\ap} -
  \frac1{2\ln2}\right).
\end{align*}
Apart from the discrepancy by $2\ln2$ which appears in the BSFT
calculations of the tachyon mass, this is consistent with the familiar
result that as a parallel $Dp$-brane and $Dp$-anti-brane are moved
toward each other, the lowest open string scalar mode becomes
tachyonic at separations $|\varphi|^2<2\pi^2\ap$ \cite{Banks:1995ch}.
We see that the separated $(\DD)_p$ system, although classically
stable, is quantum mechanically unstable for $|\varphi|^2 >
\frac{2\pi^2\ap}{2\ln2}$, with a tunneling barrier which increases
with their separation, as in Figure~\ref{figure_potential}. The system
remains unstable at the critical separation $|\varphi_c|^2 \equiv
\frac{2\pi^2\ap}{2\ln2}$ (the dotted blue line in
Figure~\ref{figure_potential}), since the $|T|^4$ term in the
potential has a negative coefficient there.  This potential was first
written down by Hashimoto \cite{Hashimoto:2002xt} assuming a linear
tachyon profile; here we have justified its form for arbitrary $T$,
which allows us to calculate the instanton ``bounce'', which is
spherically symmetric in $p+1$ dimensional Euclidean space.

That the separated $(\DD)_p$ system will annihilate via quantum
mechanical tunnelling has been studied in the literature
\cite{Callan:1998kz, Hashimoto:2002xt}.  Here we shall use the above
effective action to find the decay rate and check the validity of the
thin wall approximation.  For fixed brane separation (that is,
constant $|\varphi|$, a very good approximation in the slow-roll phase
during the inflationary epoch in the early universe), we calculate the
decay rate.  We aim to do so including the contribution from all
kinetic terms.  The calculation is tractable since the tachyon decays
in one direction in field space only, $T = \bar T$.  We set the value
of the tachyon potential at the false vacuum, $T=0$ to be zero, and
the resulting Euclidean Lagrangian for the bounce \cite{Coleman:1985}
becomes
\begin{align*}
  &\mathcal L_E = 2\tau_p \sqrt{g_E}\;\left[e^{-2\pi\ap T^2}
  \F(\frac1\pi|\varphi|^2T^2)
  \F(4\pi\ap^2\partial_\mu T\partial^\mu T) - 1\right],
  &\tau_p = \frac1{(2\pi)^pg_s\ap^{\frac{p+1}2}}.
\end{align*}
The tunneling rate can be computed numerically by the standard
instanton methods, where the probability of tunneling is
\begin{align*}
  \mathcal{P} \sim K(\varphi) e^{-S_E(\varphi)},
\end{align*}
$S_E(\varphi)$ being the Euclidean action of the instanton and the
factor $K(\varphi)$ is due to both the quantum fluctuations about the
instanton transition and to solutions of higher action which shall in
general depend on the separation, $\varphi$.  Calculating the
``bounce'' solution and integrating it numerically gives, to a good
approximation,
\begin{align}\label{instanton}
  &S_E(\varphi) \simeq 4\pi c_1c_2^{p+1}\left[\frac{p^p}{(p+1)g_s}
  \frac{2\pi^{\frac{p+1}2}}{\Gamma\left(\frac{p+1}2\right)}\right]
  \left(\frac{|\varphi|-|\varphi_c|}{\sqrt\ap}\right)^{\frac{p+1}2},
  &\begin{array}{l}
    c_1 \sim 1.5,\\c_2 \sim 0.29,
  \end{array}
\end{align}
when $|\varphi|>|\varphi_c|$.  We have expressed $S_E$ in this form to
most easily compare to the expression for the thin wall approximation
\cite{Coleman:1985},
\begin{align*}
  S_E(\varphi) \simeq \left[\frac{p^p}{(p+1)}
    \frac{2\pi^{\frac{p+1}2}}{\Gamma\left(\frac{p+1}2\right)}\right] 
  \left(\frac{S_1}{\epsilon_p}\right)^{p+1}\epsilon_p
\end{align*}
where $S_1$ is the action for the one-dimensional instanton and 
$\epsilon_p = 2\tau_p$.  This imples that the thin wall bounce has the
form
\begin{align*}
  \frac{S_1}{\sqrt{\ap}\epsilon_p} 
  = 2\pi c_2\left(\frac{|\varphi|-|\varphi_c|}{\sqrt\ap}\right)^{\hf}.
\end{align*}
In the thin wall approximation $c_1=1$, and comparison with the
numerical result shows that $c_2 \simeq 0.29$.  We expect the thin
wall approximation to be valid when $\varphi$ becomes large. Note that
$S_1$ differs from that in \cite{Hashimoto:2002xt} (in which $S_1$ is
linear in $\varphi$) because that calculation was performed with
truncated kinetic and potential terms.

Classically, for large enough separation, when $m_T^2>0$, the ground
state is $T=0$, and $V(0,\varphi)=1$.  This implies that there is no
force in the $(\DD)_p$ system.  However, since the system is
non-supersymmetric, quantum corrections are clearly present.  It is
known that the one-loop open string contribution is dual to the closed
string exchange.  For large separation, this is dominated by the
exchanges of the graviton, dilaton and RR field $C_{p+1}$ between the
D$p$ and the anti D$p$-brane and has been calculated.  These one-loop
open string corrections can be included by inserting the classical
closed string background produced by a D$p$ and a $\bar{\text{D}}p$
brane into the $(\DD)_p$ action.  The supergravity solution is well
known \cite{Horowitz:1991cd}.
\begin{align*}
  ds^2 &= \h{\hf}\left(-dt^2+\sum_{i=1}^pdx^idx^i\right)
  + \h{-\frac32-\frac{5-p}{7-p}}dr^2 
  + r^2\h{\hf-\frac{5-p}{7-p}}d\Omega_{8-p}^2,\\
  e^{-2\phi} &= g_s^{-2}\h{-\frac{p-3}2},
  \quad\quad\quad h(r) = 1 - \frac{g_s\beta}{r^{7-p}}\\
  (C_{p+1})_{a_1\ldots a_{p+1}} &= \frac{\beta}{r^{7-p}}
  \epsilon_{a_1\ldots a_{p+1}},
  \quad\quad\quad \beta \equiv (4\pi)^{\frac{5-p}2}\ap^{\frac{7-p}2} 
  \Gamma\left(\frac{7-p}2\right).
\end{align*}
This classical closed string background of a brane shall be ``felt''
by the anti-brane, so we insert this into that part of the action
corresponding to an anti-brane and the similar background into the
brane action.  On the brane worldvolumes the separation, $r$, is
represented by the scalar field $|\varphi|$.  The result of performing
these steps is that when $|\varphi|^2\gg\ap$, the total tension of the
system is renormalized:
\begin{align*}
  S &= S_{\DD}(\varphi) + S_{\text{CS}}(\varphi),\\
  \tau_p &\to \tau_p(\varphi) = \tau_p\Bigg(
  \overbrace{1-\frac{g_s\beta}{|\varphi|^{7-p}}}^{\text{NS-NS}}
  \underbrace{-\frac{g_s\beta}{|\varphi|^{7-p}}}_{\text{RR}}\Bigg).
\end{align*}
Clearly for a brane-brane system, the sign of the RR contribution is
reversed and the tension is unrenormalized.  The renormalized tension
then gives a potential for the scalar representing the separation, and
we see there is an attractive force between the brane and anti-brane.
When the brane separation decreases, massive closed string modes start
to contribute to $\tau_p(\varphi)$. Their contributions are easy to
include, except when the brane separation becomes so small that
$m_T^2$ becomes negative. When the tachyon appears, $\tau_p(\varphi)$
becomes complex.  Fortunately, $\tau_p(\varphi)$ is expected to remain
finite \cite{Garcia-Bellido:2001ky} and tachyon rolling happens so
fast that the precise form of $\tau_p(\varphi)$ at short distance
becomes phenomenologically unimportant \cite{Shiu:2002xp}.

\section{Conclusion}
In this paper, we present a fully covariant $\DD$ action
(Eq.(\ref{action})) based on boundary superstring field theory.  The
kinetic term has some rather novel features.  Its form is almost
completely dictated by BSFT and symmetry properties of the system.  It
is quite amazing that exact multi-vortex multi-anti-vortex solutions
(with arbitrary positions and arbitrary constant velocities) can be
easily written down (Eq.(\ref{general_solution})).  The simplicity of
these analytic solutions may be very useful in the study of the
production of vortices.  In the early universe in brane world
scenarios, the production of such vortices corresponds to the
production of cosmic strings towards the end of the inflationary
epoch.
 
The solitonic solution has a large peculiar degeneracy: the energy and
the RR charge of the solutions depend only on the vorticities and not
on the ``magnetic'' flux that may or may not be present inside the
vortices.  Further improvement on the $\DD$ action should lift this
degeneracy.  However we are unable to answer the question that we
endeavor to address: as a soliton in the exact theory, whether the
D$p$-brane has a ``magnetic'' flux in its core.  If the degeneracy is
lifted at the classical level, one should simply go back to the path
integral expression (\ref{definition_S}) for the $\DD$ action, which
is supposedly classically exact, to reexamine the solitonic solutions.

The inclusion of the coupling of the gauge field to the tachyon in the
$(\DD)_p$ action allows us, via T-duality, to consider the situation
when the brane and the anti-brane are separated. This action is
suitable for the study of the inflationary scenario in the brane
world. During the inflationary epoch in early universe, the branes
move slowly towards each other, since the probability of their
annihilation through tunneling is exponentially small.  Toward the end
of inflation, the annihilation process described by tachyon
condensation should be accompanied by the reheating of the universe,
the defect production and tachyon matter production.  The $(\DD)_p$
action and its generalization should provide a firm framework to study
these phenomena.

\acknowledgments
We thank Koji Hashimoto, Ra\'ul Rabad\'an, Vatche Sahakian, Ashoke
Sen, Samson Shatashvili, Gary Shiu, Horace Stoica and Ira Wasserman
for useful discussions. This material is based upon work supported by
the National Science Foundation under Grant No.~PHY-0098631.

\providecommand{\href}[2]{#2}\begingroup\raggedright
\endgroup

\end{document}